# The Methods for Direct Detection of WIMP with Mass ≤ 0.5 GeV

B. M. Ovchinnikov[*], I. I. Tkachev, V. V. Parusov

Institute for Nuclear Research, Russian Academy of Sciences, Moscow, Russia

**Abstract**

The chamber for direct detection of WIMP with mass ≤ 0.5 Gev was developed. The chamber is filled with gas mixture Ne+10% Hydrogen +0,15ppm Tetramethylgermanium (TMG). For events detections used GEM+pin-anodes, which provide the energy threshold about eV. The electron background is suppressed owing to photosensitive addition TMG. It is proposed also for direct detection of WIMP the liquid argon chamber with Hydrogen dissolved in liquid argon at a concentration 100ppm+0,015ppm TMG.

## 1. Introduction

The measurement of cross section of WIMP scattering on proton is necessary to clear up the dark matter nature [1, 2].

Astronomical observation give strong avidense for existence of non-luminous and nonbarionic matter, presumably composed of a new tipe of elementary particles. The detectors with pure NaI [3], Xe or Ar allow to search the WIMP with large mass (of dozens or hundreds GeV), because the energy of nuclear recoils in these detectors from low mass WIMP are low. The experiments with nobles gases are shown in Table1. To account for yearly modulation effect in DAMA-LIBRA experiment [3] J.Va'vra [4] have supposed that this effect is explained by low mass WIMP scattering on protons in $H_2O$ molecules which contamination about 1ppm in NaI crystals.

## 2. Spherical Proportional Detector

The spherical proportional detector was developed for search the low mass WIMP [5].

This detector was filled with $H_2$ or Ne and has the energy threshold about 100 eV.

## 3. Double-Phase Argon Chamber

The double-phase argon chamber with mass up to $10^4$ tons was proposed for WIMP detection in our work [6]. For electron background suppression was proposed the photosensitive addition Ge $(CH_3)_4$ [7].

For detecting events in gas-phase was developed the system metallic GEM[8]+ pin –anode with 10%$H_2$ addition and $K_{ampl}$=5·$10^7$ [9]. The concentration $H_2$ in liquid Ar is equal about 100 ppm, this allows to detect the low mass WIMP (≤0.5GeV) also because the concentration $H_2$ is 100 time more then in [3].

By comparing the work[5] ,where the energy threshold is equal ~100 eV ,with amplification factor of detecting system~$10^4$, we can to estimate the threshold of our experiment as 100 eV·$10^4$/5·$10^7$ ~1eV.

\* Corresponding author
E-mail address: ovchin@inr.ru (B. M. Ovchinnikov)

## 4. The Chamber with Ne+10%H$_2$(0-1bar) Filling

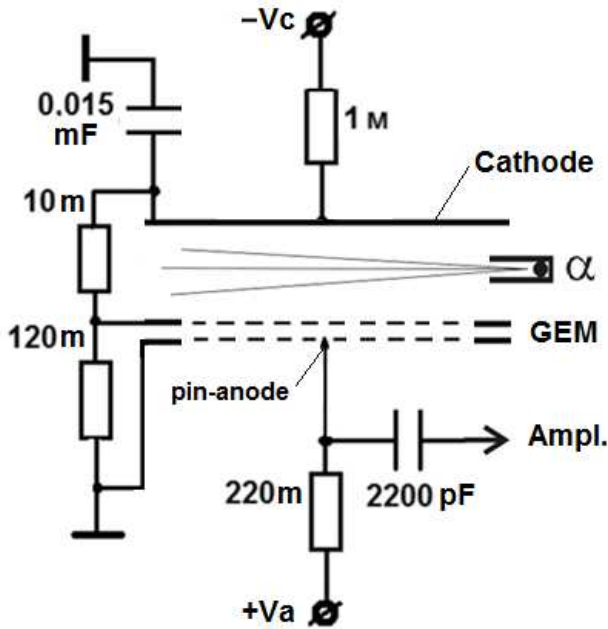

**Figure 1.** GEM+pin-anode.

On Fig. 1 is shown the system GEM+ pin-anode, which is used for events detecting in this chamber and in double-phase argon chamber. The front of signal in this system is equal ≤3 μsec.The detecting of front allows to measure the event dimension in z-direction for electron background suppression[5].

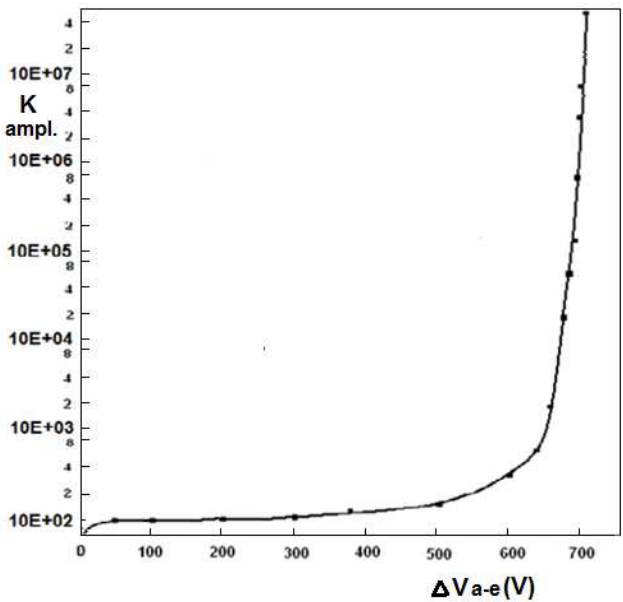

**Figure 2.** The dependence of K$_{ampl.}$ on potential difference between the pin-anode and GEM, K$_{ampl}$= Q$_{ampl}$/Q$_{ioniz}$ ,where Q$_{ampl}$ –the charge detected, Q$_{ioniz}$ –the ionization charge.

For purification of gases are used the methods with Ni/SiO$_2$ adsorbents[10].

The addition in chamber of Ge (CH$_3$)$_4$ allows to suppress the electron background (gamma-background , Ar$^{39}$ and tritium decays).

On Fig. 2 the dependence of K$_{ampl.}$ on potential difference between the pin-anode and GEM is shown .The use in chamber of spectrometric amplifier allows to obtain the energy threshold about eV. This energy threshold allows to search the WIMP with mass ≤0.5 GeV/c$^2$ (see table2).Double-phase Ar chamber or the chamber with Ne+10%H$_2$ filling are placed in low background laboratory in low background shielding for search the yearly or daily modulation effects.

The double-phase argon chamber and chamber with Ne-filling allow to search the axions, emitted from the sun.

The chamber with Ne+10%H$_2$ filling is shown on Fig.3.

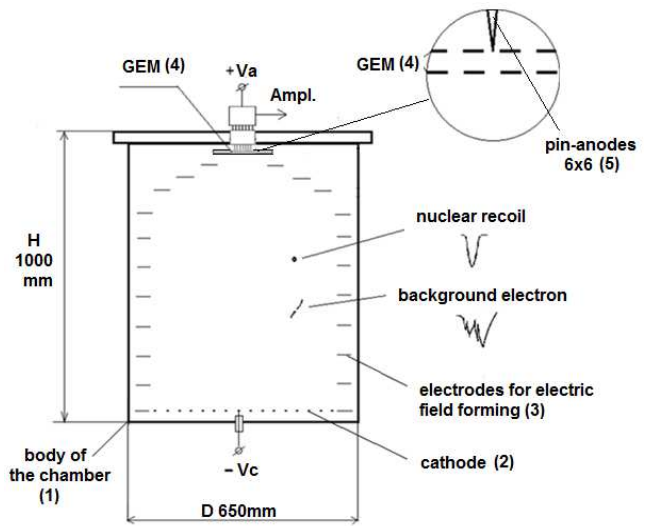

**Figure 3.** The chamber with Ne+10%H$_2$ filling.

The chamber consists of titanium body 1,of detecting system (metallic GEM 4 +36 pin-anodes 5),of wire cathode 2,winded with 0,1 mm in diameter of beryllium bronze wires. The positive voltage on cathode with respect to chamber body suppress the background from cathode. The electrodes 3, produced from pure copper ,forms in volume of the chamber the electric field.

For improvement sensitivity of chamber to low mass WIMP it is necessary to add in Ne-filling of the chamber the 1ppm TMAE (tetrakis dimethylamino ethylene) [7], which has the low ionization potential 5.36eV.When the recoil proton come into collision with TMAE molecule, the molecule TMAE is ionized and the signal from low mass WIMP is increased.

**Table 1.** The detectors for direct DM searches.

| The name of project | The target of detector | The detection method | The threshold of detection | The method for background suppression | $Ar^{39}$ concentration in Ar | The expected result |
|---|---|---|---|---|---|---|
| "ArDM" A.Rubbia [11] | Ar 1000(850) kg double-phase | $S_2/S_1$+F PMT+GEM | $E_{nr}^{min}$=30 кэВ | $S_2/S_1$+F* | $10^2$ decays/t·s | σ(WIMP) =$10^{-45}$cm² |
| "MiniClean" Los Alamos [12] | Ar liquid 500(150)kg single-phase | F 92 PMT | $E_{nr}^{min}$=30кэВ $M^{min}$(WIMP) =20 GeV | F | $10^3$ decay/t·s | $10^{-45}$см² |
| "Deap-3600" Los Alamos [12] | Ar liquid 3600(1000)kg single-phase | F 266 PMT |  | F | $10^3$ decays/t·s | $10^{-46}$ cm² |
| "Clean" Los Alamos [12] | Ar liquid 40(10)tons single-phase | F PMT | $M^{min}$(WIMP)=60 GeV | F | <$10^2$ decays/t·s | 6·$10^{-47}$ cm² |
| "Darvin" [13] | Ar 20(10)tons double-phase | $S_2/S_1$+F avalanche photodiodes +GEM | $E_{nr}^{min}$=30 кэВ | $S_2/S_1$+FK suppression=$10^8$ | <40 mBq/kg | 4·$10^{-48}$ cm² |
|  | Xe 8(5) tons double-phase | F+$S_2/S_1$ avalanche photodiodes +GEM | $E_{nr}^{min}$=10 кэВ | F+$S_2/S_1$ K=$10^4$ | Background $10^{-4}$ decays/kg·day·keV |  |
| Los Angeles Dr.D.Cline [14] (proposal) | Ar 580(500)tons double-phase | $S_2/S_1$ 12000 avalanche Photodiodes | $M^{min}$(WIMP) ≅20GeV | $S_2/S_1$+F | <10 decays/t·s | $10^{-48}$ cm² |
|  | Xe 146(100)tons double-phase | $S_2/S_1$+F 3740 avalanche Photodiodes | $M^{min}$(WIMP) ≅6 GeV | $S_2/S_1$+F K=$10^3$ |  |  |

*For suppression of the electron background in some works (Table1) the criterion F used:
F= $I_s/I_s+I_t$ ,
$I_s$ – singlet intensities, $I_t$-triplet intensities. $S_1$-scintillation signal , $S_2$-ionization signal.

**Table 2.** Maximum calculated nuclear recoil energy $E_{keVnr}$ as a function of WIMP mass for two targets: hydrogen and sodium (Na, Ne) [3].

| WIMP mass [GeV/c2] | Nucleus | EkeVnr[keV] |
|---|---|---|
| 0.5 | H | 1.91 |
| 1.0 | H | 4.30 |
| 1.5 | H | 6.20 |
| 2.0 | H | 7.65 |
| 2.5 | H | 8.78 |
| 3.0 | H | 9.68 |
| 0.5 | Na | 0.19 |
| 1.0 | Na | 0.73 |
| 1.5 | Na | 1.57 |
| 4.0 | Na | 9.07 |

# 5. Conclusion

The method of $H_2$ addition in liquid Ar and method of event detecting, proposed in this paper allows to search the low mass WIMP in all experiments with Ar chamber.

# References

[1] V.V Titkova, V.A.Bednyakov, "Neutralino-nucleon cross section for detection of low-mass WIMP",preprint Dubna E4-2004-131.

[2] V.A. Bednyakov, B.M. Ovchinnikov, V.V. Parusov, "Search for spin-dependent interaction of Dark Matter particles", preprint INR №1144/2005.

[3] R. Bernabei, P. Belli, F. Cappella, V. Caracciolo, S. Castellano, R. Cerulli, C.J. Dai and A. d' Angelo," Final model independent result of DAMA/LIBRA-phase1", Eur. Phys. J. C73 (2013) 12, 2648R.

[4] J. Va'vra,"A New Possible Way to Explain the DAMA Results" arXiv:1401.0698v5, (2014), Physics Letters B 735 (2014)181.

[5] Y. Giomataris, I. Irastorza, I. Savvidis et al., "A Novel Large-volume Spherical Detector with Proportional Amplification read-out", JINST 3:P09007, (2008).

[6] B.M. Ovchinnikov, Yu. B. Ovchinnikov, V.V. Parusov, "Massive liquid *Ar* and *Xe* detectors for direct DM searches", JETP Lett., 96, (2012) 149-152, Universal Journal of Physics and Application 1(2): 66-70,2013.

[7] D.F. Anderson," New photosensitive dopants for liquid Ar", NIM A245 (1986) 361.

[8] D.S. Kosolapov, B.M. Ovchinnikov, V.V. Parusov, "The gas electron amplifier with metallic electrodes", Instruments and Experimental Techniques, No. 1 (2014)77.

[9] B.M. Ovchinnikov, V.V. Parusov, "Metods for Detecting Events in Double-Phase Argon Chambers", Instruments and Experimental Techniques, 2013, Vol.56, No. 5 , pp. 516-520.

[10] A.S. Barabash, O.V. Kazachenko, A.A. Golubev, B.M. Ovchinnikov, "Catalytic and adsorbing purification of NG, $H_2$, $CH_4$ from $O_2$", J. "Chemical industry", 1984, No. 6, p.373.

[11] A. Marchionni, C. Amser, A. Badertscher et al., "Ar DM: A ton-scale liquid Ar detector for direct dark matter search", arXiv:1012.5967.

[12] MiniClean Collaboration Los Alamos, "The MiniClean Dark Matter Experiment", Proceedings of the DPF-2011Conference, Providence, RI, August 8-13, 2011.

[13] Laura Baudis, "Darvin: dark matter WIMP search with noble liquids", arXiv: 1012.4764V1 [astro-ph. IM] 21 Dec. 2010.

[14] K. Arisaka, C.W. Lam, P.F. Smith at al., "Studies of a three-stage dark matter and neutrino observatory based on multi-ton combinations of liquid Xe and liquid Ar detectors", arXiv: 1107.1295V1 [astro-ph. Im] Jul. 2011.